\documentclass[%
 preprint, 
superscriptaddress,
 amsmath,amssymb,
 aps, physrev,
]{revtex4-2}

\usepackage{graphicx}
\usepackage{dcolumn}
\usepackage{bm}

\begin{document}


\title{\textbf{On the optimisation of the geometric pattern for  structured illumination based X-ray phase contrast and dark field imaging: A simulation study and its experimental validation.} 
}%

\author{Clara Magnin}
 \affiliation{Univ Grenoble Alpes, Inserm UA7, Strobe,  Grenoble, France}
 \affiliation{Xenocs SAS, Grenoble France}
 \author{Laurene Quénot}
 \affiliation{Univ Grenoble Alpes, Inserm UA7, Strobe,  Grenoble, France}

 \author{Dan Mihai Cenda}
 \affiliation{Xenocs SAS, Grenoble France}

 \author{Blandine Lantz}
 \affiliation{Xenocs SAS, Grenoble France}

 \author{Bertrand Faure}
 \affiliation{Xenocs SAS, Grenoble France}
 
\author{Emmanuel Brun}%
 \email{Contact author: emmanuel.Brun@inserm.fr}
\affiliation{%
Univ Grenoble Alpes, Inserm UA7, Strobe,  Grenoble, France}

\date{\today}

\begin{abstract}
Phase-contrast and dark-field imaging are relatively new X-ray imaging modalities that provide additional information to conventional attenuation-based imaging. However, this new information comes at the price of a more complex acquisition scheme and optical components. Among the different techniques available, such as Grating Interferometry or Edge Illumination, modulation-based and more generally single-mask/grid imaging techniques simplify these new procedures to obtain phase and dark-field images by shifting the experimental complexity to the numerical post-processing side. This family of techniques involves inserting a membrane into the X-ray beam that locally modulating the intensity to create a pattern on the detector which serves as a reference. 
 However, the topological nature of the mask used seems to determine the quality of the reconstructed phase and dark-field images. We present in this article an in-depth study of the impact of the membrane parameters used in a single mask imaging approach. A spiral topology seems to be an optimum both in terms of resolution and contrast-to-noise ratio compared to random and regular patterns. 
\end{abstract}

\maketitle


\section{Introduction}
X-ray phase contrast and dark-field imaging have emerged as powerful tools in a wide array of applications, demonstrating their efficacy in both bio-medical \cite{willer2021x,viermetz2022dark}, cultural heritage \cite{mocella2015revealing} and industrial fields \cite{partridge2022enhanced,shirani20234d}. These techniques have proven to be remarkably effective in enhancing image quality both in terms of contrast and spatial resolution as they were able to provide valuable information beyond conventional absorption-based imaging. However, their implementation outside synchrotron facilities necessitates optical setups \cite{quenot2022x}. The most common technique to transfer phase contrast on conventional X-ray source is Grating interferometry which introduces generally 2 grating interferometers that reduce the flux or augment the radiation dose to the sample and that need to be carefully aligned, generating therefore problems for their 3D implementation in rotation gantry systems \cite{viermetz2022dark}.

Over the past decade a new family of methodologies, named single mask methods \cite{morgan2012x,berujon2012two,lim2015experimental} has emerged. This set of techniques significantly simplified the X-ray phase contrast imaging (PCI) by shifting the experimental complexity towards digital and data processing realms. Single mask techniques therefore made phase contrast and dark-field techniques more accessible and practical to a wider community. Indeed, by analyzing the distortion of the original pattern caused by the introduction of a sample, these techniques allow to directly measure refraction and scattering. Different implementations exist. The modulation pattern might differ in nature (1D, 2D) and the space in which the phase retrieval is performed: either Fourier or real one. In this study, we use the concept of Modulation Based Imaging (MoBI) \cite{quenot2022Optica} which is the most generic method as no assumption is made on the modulation generated by the mask and where the phase retrieval is made in the real space.  

While these advancements have made X-ray phase contrast and dark-field imaging more feasible using conventional X-ray sources \cite{zanette2014speckle}, the optimization of modulation topology remains an open research question. A subfamily of these single grid methods uses regular patterns, either cartesian grids\cite{sun2019propagation} or honeycomb topology\cite{busi2023multi}. They profit of this known modulation to either model the signal in the real space \cite{croughan2023directional} or use Fourier space-based method \cite{sun2022grating} to simply retrieve dark-field and phase contrast. In the case of the MoBI method \cite{magnin2023dark} which derives from the synchrotron's native speckle techniques \cite{zdora2017x,berujon2016x,morgan2012x} no assumption is made of the modulation. The method consists in inserting a membrane into the beam, usually a piece of sandpaper, and making several pairs of images of this membrane with and without the presence of the sample. The membrane is moved a few pixels between each pair of images and the grains on the surface of the paper modulate the X-ray intensity randomly on the detector. 

In this study, the impact of different geometries on image quality was investigated. This work enables the optimization of modulation geometries according to the characteristics of the devices used. We aim to provide a more streamlined approach to X-ray phase contrast and dark-field imaging, making these techniques even more valuable and versatile for a wide range of applications.
We present the results obtained by simulating an existing device and the associated experimental validation. 

\section{Material and method}

\subsection{Modulation Based Imaging method}
\begin{figure}[h!]
\centering
\includegraphics[scale=0.3]{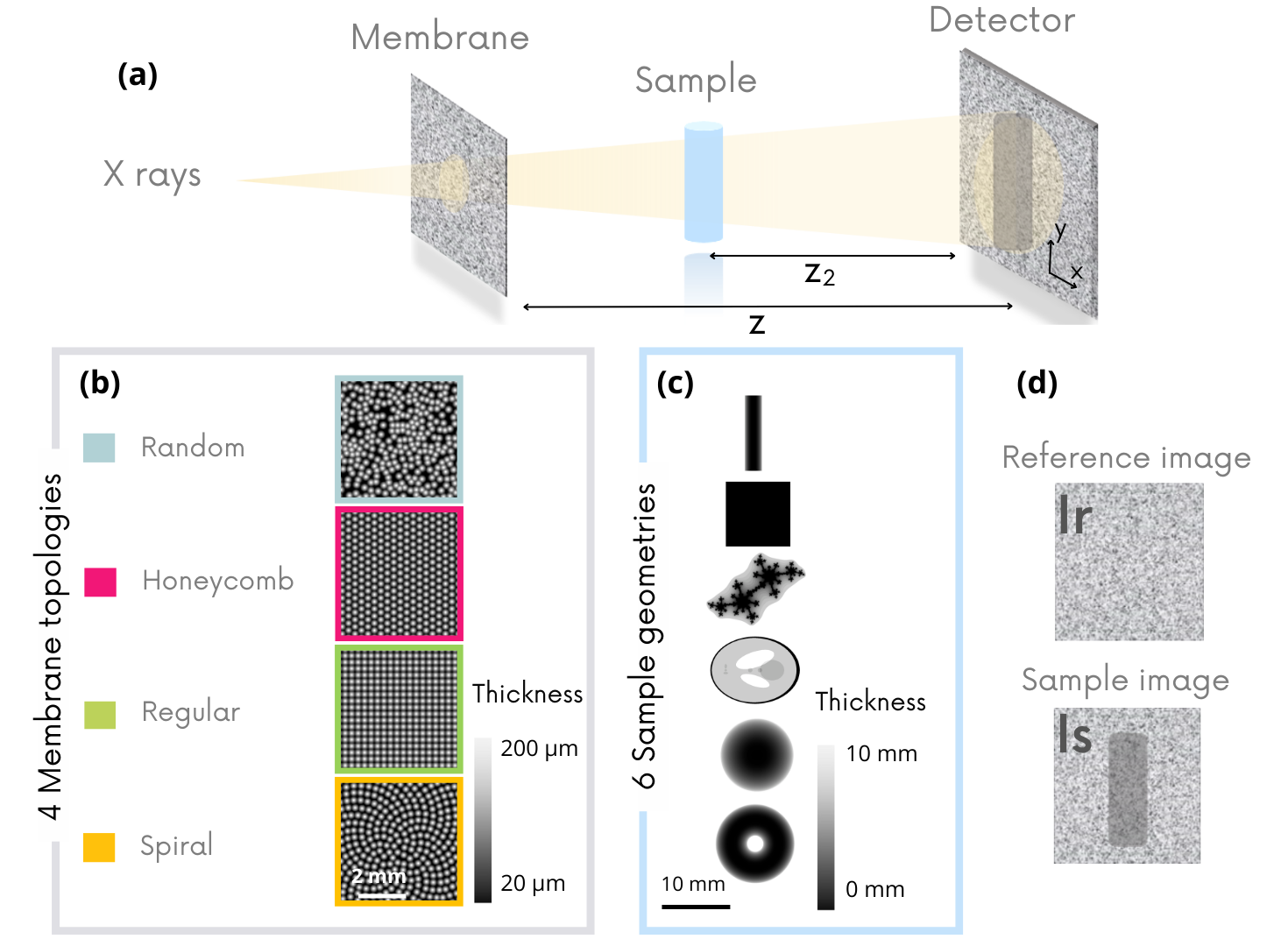}
\caption{Modulation Based imaging (MoBI) method consists in inserting a single mask into the X-ray beam to locally modulate the intensity on the detector. Several pairs of images are acquired of the mask alone (Ir) and of the mask with the sample (Is) by shifting the mask position between each acquisition pair. 
Representation of the different mask (b) and samples (c) geometries simulated}
\label{setup}
\end{figure}

In this study, the Modulation Based Imaging method (MoBI) is used to acquire and retrieve phase and dark-field images. MoBI involves inserting a membrane into the X-ray beam to modulate its intensity locally, thereby generating a reference pattern in the detector plane. In the original approach \cite{morgan2012x,berujon2012two}, authors used sandpapers to create a speckle-like pattern thanks to the coherence of synchrotron beams and the multiple refraction caused by different layers of $Si$ grains. More recently in \cite{quenot2022Optica} it was proposed to use a higher-Z element to create a modulation pattern based on absorption using a low coherence source. Nevertheless, in all these experimental setups, several pairs of images of the reference pattern alone ($I_r$) and with the sample ($I_s$) are taken. The membrane is moved of few pixels between each acquisition to obtain an entirely different reference pattern for each pair of images. Figure \ref{setup} illustrates the MoBI method and shows the different membrane and sample geometries used in the study.

In this study, 10 pairs of images of the reference alone ($I_r$) and the reference with sample ($I_s$) are used to retrieve phase and dark-field images both experimentally and by simulation. Between each pair of images, the membrane is moved by a few pixels in the 1 (x) or 2 (x and y) directions in space.
These pairs of $I_r$ and $I_s$ images will be used to retrieve phase and dark-field images after an algorithmic analysis described in section \ref{LCS}.

\subsubsection{Simulations}
To compare a wide variety of membrane types and sample geometries, a large part of this study was carried out by numerical simulation. The PARESIS simulation tool \cite{Paresis} is used to digitally generate pairs of reference and sample images, taking into account several inputs to define the geometries and parameters of the imaging device (source-detector, distances) as well as the membrane and sample characteristics.

Four different membrane types and 6 samples were numerically simulated for this purpose and are shown in Figure \ref{setup} (c). In the simulations, the samples are made of nylon with different geometries and thicknesses. Six samples were generated:
\begin{itemize}
\item A cylindrical wire with a diameter of 1.5mm
\item A sphere with a diameter of 12 mm
\item A cube with side length of 7 mm
\item A torus with an outer diameter of 13 mm, an inside diameter of 3mm and a maximum thickness of 3 mm
\item A Shepp-Logan phantom of 13*18 mm with thicknesses ranging from 1 to 5 mm.
\item A Julia's fractal geometry of 18*10 mm with a maximum height of 5 mm
\end{itemize}

The samples geometries was chosen to represent simple and more complex materials. The Shepp-Logan (SL) phantom was used to assess contrast-to-noise ratio and quantitativeness in a more complex geometry. Finally , the fractal sample was mainly used to assess the quality of reconstructed images; its geometry is complex with fine details enabling the reconstruction to be probed on several size scales.

In parallel, membranes are created digitally. Their construction is inspired by the sandpaper used to date, where sand grains on the surface of the paper allow variations in X-ray attenuation. To reproduce similar intensity gradients, $SiO_2$ cones with a fixed height of 200 nm and variable diameters are digitally placed on the surface of the membranes. These cones are distributed in different arrangements over a flat surface of the size of the detector. 
Four families of membrane geometries have been numerically created and are shown in Figure \ref{setup} (b):
\begin{itemize}
\item A random geometry reproducing the distribution of grains on the surface of sandpapers
\item A hexagonal geometry that mimics honeycomb constructions known to optimize space
\item A regular geometry based on a square-mesh grid
\item A spiral geometry. 
\end{itemize}

Generally speaking, a spiral is a curve that wraps around a fixed point while moving progressively away from it. Mathematically, a spiral can be described in a polar coordinate system by an equation of the form :
$$r = f(\theta)$$
with $r$ the distance from the point to the center, $\theta$ the angle in radians and $f(\theta)$ the function that defines how $r$ varies with angle.
In this study a Vogel spiral is used, a pattern known for reproducing the geometry of sunflowers. Several spiral geometries exist, and a numerical comparison of the performance comparison of different patterns is given in the Supplement. 

Because we want to validate the results obtained by simulation on an existing system, the simulated experimental set-up is based on a Xeuss 3.0 instrument (Xenocs SAS, Grenoble France) adapted for MoBI. In this context, the simulations were carried out using the following parameters: 
\begin{itemize}
    \item Source to membrane distance : 0.33 m
    \item Source to sample distance : 0.55 m
    \item Sample to detector: 0.800 m
    \item Simulated source : Cu target with a polychromatic beam of mean energy equal to 8.6 keV
    \item Constant mean number of photons on the detector: 75000 counts
    \item Under vacuum environment
    \item Photon Counting detector of 1030*514 pixels with a pixel size of 75 µm
    
\end{itemize}

\subsubsection{Experimental setup}
As previously said, our experimental setup is a Xeuss 3.0 (Xenocs SAS, Grenoble France) adapted for MoBI. The device features a polychromatic cone-shaped beam X-ray source dedicated to imaging. A vacuum chamber contains the sample, the motorized membrane in one direction, and a movable detector along the X-ray propagation direction. 
Three membranes were used experimentally : a piece of sandpaper (random pattern), an honeycomb geometry and a Vogel spiral \cite{vogel1979better}. These membranes were fabricated by drilling circular apertures in the surface of Ni plates. The effect of the holes in the material, convoluted by the size of the source, enabled us to obtain intensity modulations similar to those obtained by adding material in the form of cones, as we had initially envisaged.
The experimental parameters of this set-up are summarised in Table \ref{tab:ExpPram}. In the rest of the article, we refer to pixel dimensions and not physical to be as generic as possible.

\begin{table}[h!]
\caption{Experimental setup parameters}

\newcolumntype{M}[1]{>{\raggedright}m{#1}}
\begin{tabular}{|M{2cm}|M{12cm}|}
    \hline
     Distances   & Source to sample: 0.55 m -  Source to membrane: 0.33 m - Sample to detector: 0.800 m   \tabularnewline
    \hline
     X-ray  source  & Copper anode - 30 kVp - 8.6 keV average energy - 50 $\mu$m spot size \tabularnewline
    \hline
     Detector  & Photon counting Eiger2 (DECTRIS, Switzerland) with pixels of 75 $\mu$m  - Imaging acquisition time: 30 s per image  \tabularnewline
    \hline
     Membrane  & Piece of sandpaper \textit{p180} (31 mm * 20 mm) and 2 micro perforated pattern of holes of 60 µm diameter in a 30-µm-thick Ni foil (31 mm * 20 mm) \tabularnewline
    
    \hline
\end{tabular}
\label{tab:ExpPram}
\end{table}

The samples used in the experimental part were i) a home-made phantom of nylon and bumdles of carbon fibres  and ii) the Julia fractal printed out in 3D in resin with the same geometrical characteristics as the simulated one.

\subsection{Phase retrieval}
\label{LCS}
Once the reference and sample image pairs have been obtained experimentally or by simulation, the phase and dark-field images were retrieved using the recently published implicit "Low coherence system" LCS reconstruction algorithm with (directional) dark-field extraction \cite{magnin2023dark}. This algorithm retrieves several image modalities: absorption, phase obtained by integrating displacement images $D_x$ and $D_y$, dark-field and directional dark-field. LCS maps the distortions in the reference pattern induced by the presence of the sample. These distortions of various kinds can be linked to the absorption, refraction and scattering phenomena that correspond to the transmission, phase and dark-field images respectively. The algorithm is based on the resolution of equations based on TIE intensity transport and optical flow conservation. More details can be found in \cite{magnin2023dark} but the general form of the founding equation is : 
\begin{equation}
\begin{aligned}
\label{eq:TIE_LCS}
I_r(x,y)-\frac{I_s(x,y)}{I_{obj}(x,y)}& \simeq D_\perp(x,y)\nabla_\perp[I_r(x,y)]\\&-z_2 D_f(x,y)\nabla_\perp^2[I_r(x,y)]
\end{aligned}
\end{equation}
Where $I_r$ and $I_s$ are the reference image and the sample image. $z_2$ is the sample-to-detector distance, $D_{\perp}=(D_x, D_y)$ is the transverse displacement field and $\nabla_{\perp}={\partial}/{\partial x}+{\partial}/{\partial y}$ is the two dimensional transverse gradient operator. $(D_x, D_y)$ provide the phase image after an integration step. $I_{obj}$ is a sink term introduce to compensate for attenuation that might comprise also the interference fringes, if any and $D_f$ the scattering term that represent the dark-field contribution. 
In order to retrieve the absorption, phase and dark-field signal, we must solve a system with at least 4 membrane positions due to the 4 unknowns ($I_{obj}$, $D_x$, $D_y$, $D_f$), the phase being the integral of $D_x$ and $D_y$:

\begin{equation}
\begin{aligned}
\label{LCSDFsystem}
I_{r}^{(k)}(x,y) &= \frac{1}{I_{obj}(x,y)}I_{s}^{(k)}(x,y)+  D_x(x,y)\frac{\partial I_r^{(k)}(x,y)}{\partial x}\\&+D_y(x,y)\frac{\partial I_r^{(k)}(x,y)}{\partial y}-z_2D_f(x,y)\nabla_\perp^2[I_r^{(k)}(x,y)],
\end{aligned}
\end{equation}

Note that in equation \ref{LCSDFsystem} the refraction terms ($D_x$ and $D_y$) are retrieved thanks to the gradient image of the reference pattern and that the dark-field term $D_f$ uses the Laplacian of the same image. Therefore, it seems obvious that optimizing the gradient and the laplacian of the modulation pattern would increase the global image quality.

A detailed review article on the different algorithms in MoBI can be found in \cite{celestre2024review}. 

\subsection{Metrics}
To assess the quality of the reconstructed images, several image quality assessment (IQA) indexes were used. In this study IQA are calculated on $D_y$ displacement images because they are directly linked to phase information (phase is the integral of the $D_x$ and $D_y$ images), and avoids the integration issues that can occur with phase images. Most of these indices were calculated using the Pytorch Image Quality \cite{piq} and Sk-Image \cite{sk} libraries:
\begin{itemize}
    \item \textbf{NRMSE} (Normalized Root Mean Square Error). RMSE is defined as the square root of the mean square error between two different images. It measures the difference between two images by calculating the average of the squared deviations for each pixel, then taking the square root of this average as :
\[ \text{RMSE}(I_0, I_1) = \sqrt{\frac{1}{N} \sum_{i=1}^{N} (I_0(i) - I_1(i))^2} \]

where \( I_0 \) is the reference image and \( I_1\) the image to be compared, \( i \) represents a pixel, and \( N \) is the total number of pixels in the images.
To facilitate comparisons, the RMSE is normalized by the Euclidean norm of the reference image, giving the NRMSE criterion. The lower the NRMSE the more similar the images are.

    \item \textbf{SSIM} (Structural Similarity Index Measure). To be more representative of the human vision than MSE the SSIM index has been introduced in the early 2000s \cite{SSIM}. This index is calculated between two windows x and y of size NxN of two images is : 

$${SSIM}(x,y)=l(x,y).c(x,y).s(x,y)={\frac {(2\mu _{x}\mu _{y}+c_{1})(2\sigma _{x}\sigma _{y}+c_{2})(cov_{xy}+c_{3})}{(\mu _{x}^{2}+\mu _{y}^{2}+c_{1})(\sigma _{x}^{2}+\sigma _{y}^{2}+c_{2})(\sigma _{x}\sigma _{y}+c_{3})}}$$

with $\mu _{x}$ the mean of x, $\mu _{y}$ the mean of y, $\sigma _{x}^{2}$ the variance of x, $\sigma _{y}^{2}$ the variance of y,
$cov_{xy}$ the covariance of x and y, $c_{1}=(k_{1}L)^{2}$, $c_{2}=(k_{2}L)^{2}$ and $c_{3}={c_{2} \over 2}$ three variables designed to stabilize division when the denominator is very low. L is the dynamic range of pixel values, i.e. 255 for 8-bit coded images.
$ k_{1} = 0,01$ et $k_2 = 0,03$ by default.
Several more recent image quality indexes have been developed from the SSIM model, such as the MS-SSIM (Multi-Scale Structural Similarity Index) and SR-SSIM (Spectral Residual based Similarity) mentioned below.

\item \textbf{MS-SSIM} (Multi-Scale Structural Similarity Index) \cite{MSSSIM} takes into account image details at different resolutions. This index aims to better reflect the way humans perceive image quality. It works by applying a low-pass filter to the image and halving its size at each iteration. The process is repeated several times, producing several scales of the image. At each scale, contrast and structure comparisons are calculated. The measurements obtained at each scale are then combined to produce an overall assessment of structural similarity.

\item \textbf{SR-SIM} (Spectral Residual based Similarity) \cite{SRSIM},The SRVS is derived from the image spectrum, where the “spectral residual” is obtained from the logarithmic spectrum. This residual is then transformed into the spatial domain to obtain a visual saliency map for each image, and compared with each other to obtain a local similarity measure. SRVS is effective for detecting salient areas, but less so for capturing local contrast, which is why SR-SIM also incorporates a measure of image gradient (GM). GM provides an additional measure for assessing local image quality, focusing on transitions and contours.

\item \textbf{VSI} (Visual Saliency-induced Index) is based on the visual importance of different image regions. Unlike other indices, VSI uses a model of visual salience, describe in the article \cite{VSI}, to identify the parts of the image that attract the human observer's attention the most. It thus gives greater weight to salient regions in the quality calculation, offering an assessment more faithful to human perception.

\end{itemize}
These indexes are measured on the reconstructed Dy displacement images rather than on the phase image itself, to avoid errors that may be linked to the integration stage.
The reference Dy images are computed from the geometries of the various samples in order to compare the simulated or experimental results to the theory.

\begin{itemize}
    \item \textbf{FRC} (Fourier Ring Correlation) is an image quality criterion widely used to assess image resolution in microscopy \cite{koho2019fourier}. This index measures the similarity between two images or sets of images by decomposing them into their frequency representations and then calculating the correlation of these frequency components at different spatial scales. This method allows us to assess the resolution to which fine image structures are correlated. In our case, we took two different realisations (two different sets of references and sample images) to evaluate the spatial resolution.
    The FRC curve shows how the correlation varies with spatial frequency. A threshold value is often defined to determine the resolution at which correlation becomes insignificant. This threshold value is interpreted as the resolution at which details can no longer be distinguished with confidence in the images. In this analysis, the most widely used threshold of (1/7 =0.143) is chosen.
\end{itemize}

\section{Results}

\subsection{Membrane optimization by numerical simulation}
\begin{figure}[h!]
\centering
\includegraphics[scale=0.3]{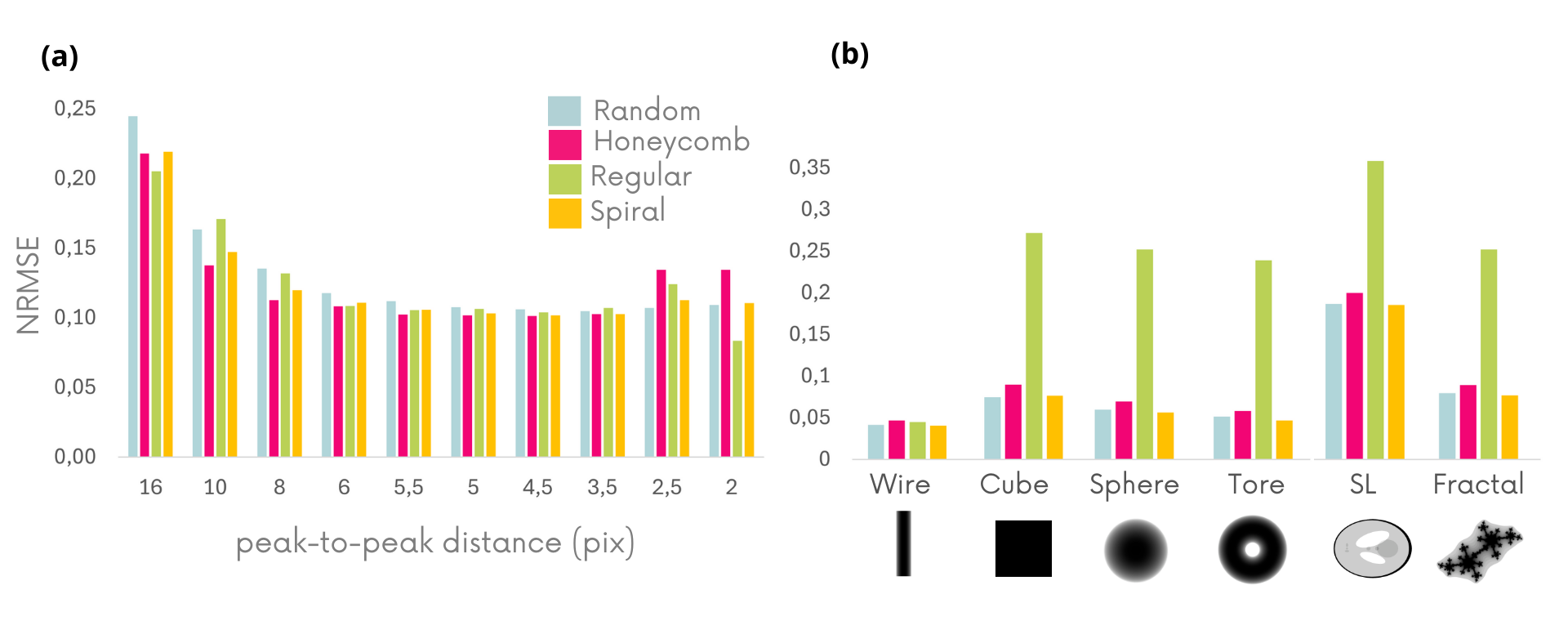}
\caption{Normalised root mean square error (NRMSE) index of the retrieved images as a function of average peak-to-peak distance between modulations on the membrane; comparing the 4 membrane geometries on the fractal sample. The membrane is moved in 2 directions (a). NRMSE index calculated on the 6 samples $D_y$ images retrieved according to different membranes. The membrane is moved in 1 direction only (b). The lower the index, the better the image quality}
\label{mod}
\end{figure}

The first step in objectively comparing the 4 membrane topologies was to set the optimal configurations in terms of gradient distribution on the reference image for each geometry. The modulation amplitude was fixed by keeping the same height of cone on our virtual membrane and we studied the influence of the modulation size on the image quality. Figure \ref{mod} (a) shows the evolution of NRMSE criterion on the $D_y$ images of a fractal object for different modulation sizes (expressed as the peak-to-peak distance). The fractal object was chosen for this purpose because it's a multi-scale geometry that allows to probe the resolution of the images. The membrane was virtually moved randomly in two spatial directions (x,y), perpendicular to the beam propagation axis, between each of the 10 reference/sample image pairs. We remind the reader that the lower the NRMSE is the better the image quality. We can observe the same plateau in the image quality measurements for all the different topologies at approximately the same modulation size: 6-3 pixels. This latter is a clear advantage of the MoBI approach because it means that the same membrane can be used for different magnifications.

To avoid to optimize for a given algorithm we also used two explicit tracking algorithms: Unified Modulation Pattern Analysis (UMPA) \cite{zdora2017x} and X-ray Speckle Vector Tracking (XSVT) \cite{berujon2016x}. Please see Supplementary material for results, briefly the maximum image quality is found for a modulation size of 6 to 3 pixels for all the algorithms, with LCS outperforming in terms of image quality. It is beyond the scope of this article to compare the different algorithms in detail, and we refer the reader to \cite{celestre2024review} for a quantitative comparison.

In the rest of this study, the results were obtained using a single-direction mask movement (for practical reasons of using a single motor). Image quality measurements made with two directions of membrane sensing can be found in Supplementary material. In the simulation section, we use 10 different positions of the membrane obtained from the same initial membrane by applying a random displacement of 50 to 150 pixels in the x direction.

The sub-figure \ref{mod} (b) shows NRMSE index measurements obtained on $D_y$ images of the different samples using the 4 membrane geometries with single-axis displacement. Under these conditions, the regular pattern produces poor-quality images due to artifacts except for the wire sample which is orthogonal to the membrane displacement. The refraction effect caused by the wire is unidirectional and can therefore be measured without being impacted by the periodicity effect of the regular membrane. The scores obtained by the 3 others membrane topologies are closer to each other. However, it is interesting to note that the random pattern, used experimentally up to now with sandpaper \cite{zanette2014speckle,berujon2012two,morgan2012x} or powder deposits \cite{quenot2022Optica}, is not the most interesting option. The spiral pattern seems to have a slight advantage over the other geometries, as it achieves the lowest NRMSE values in all cases. 

\begin{figure}[h!]
\centering
\includegraphics[scale=0.25]{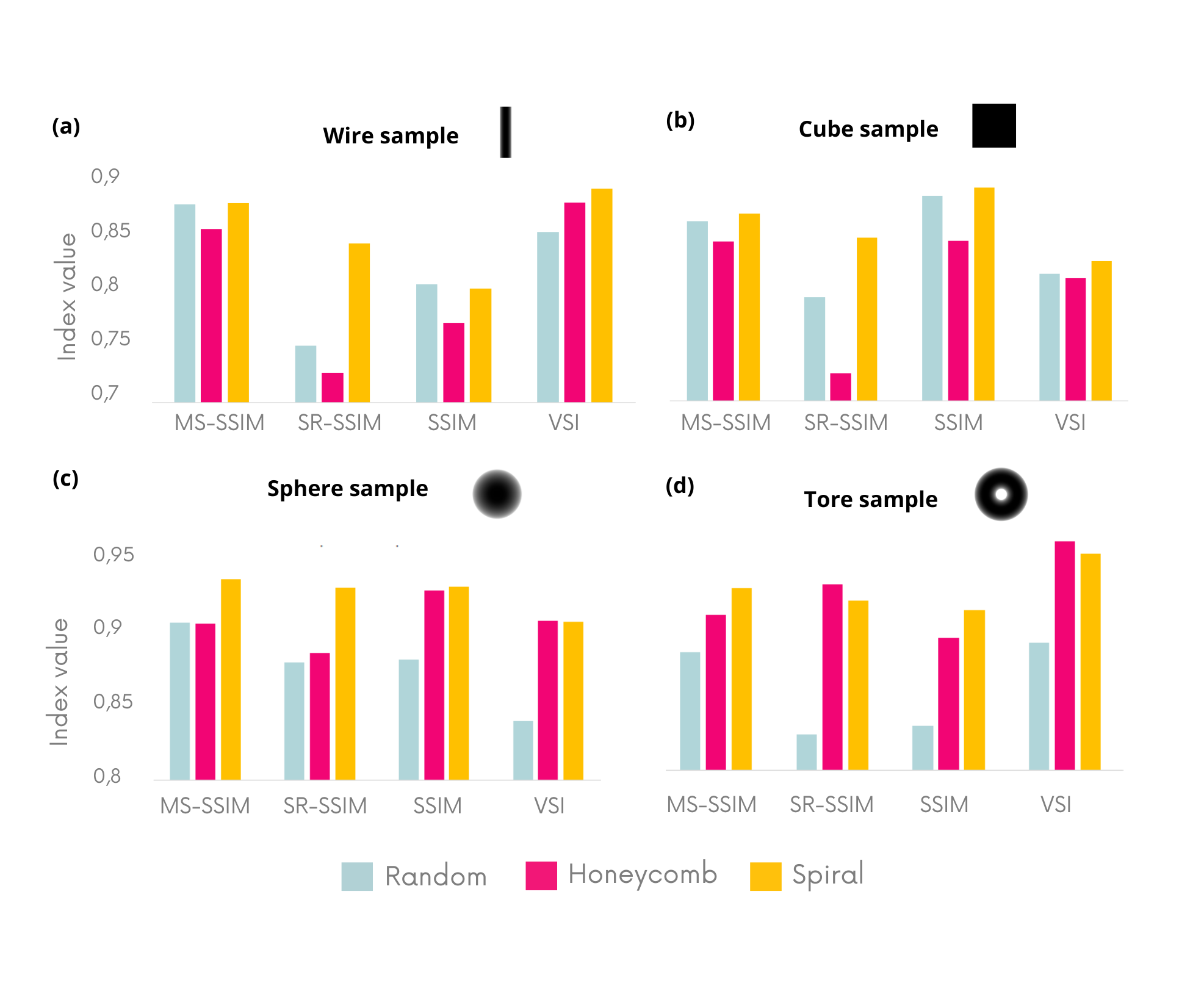}
\caption{Different image quality indexes calculated on sample $D_y$ images retrieved according to different membrane geometries. The higher the indexes, the better the image quality. The membrane is moved in one direction only.}
\label{indexes}
\end{figure}

\begin{figure}[h!]
\centering
\includegraphics[scale=0.35]{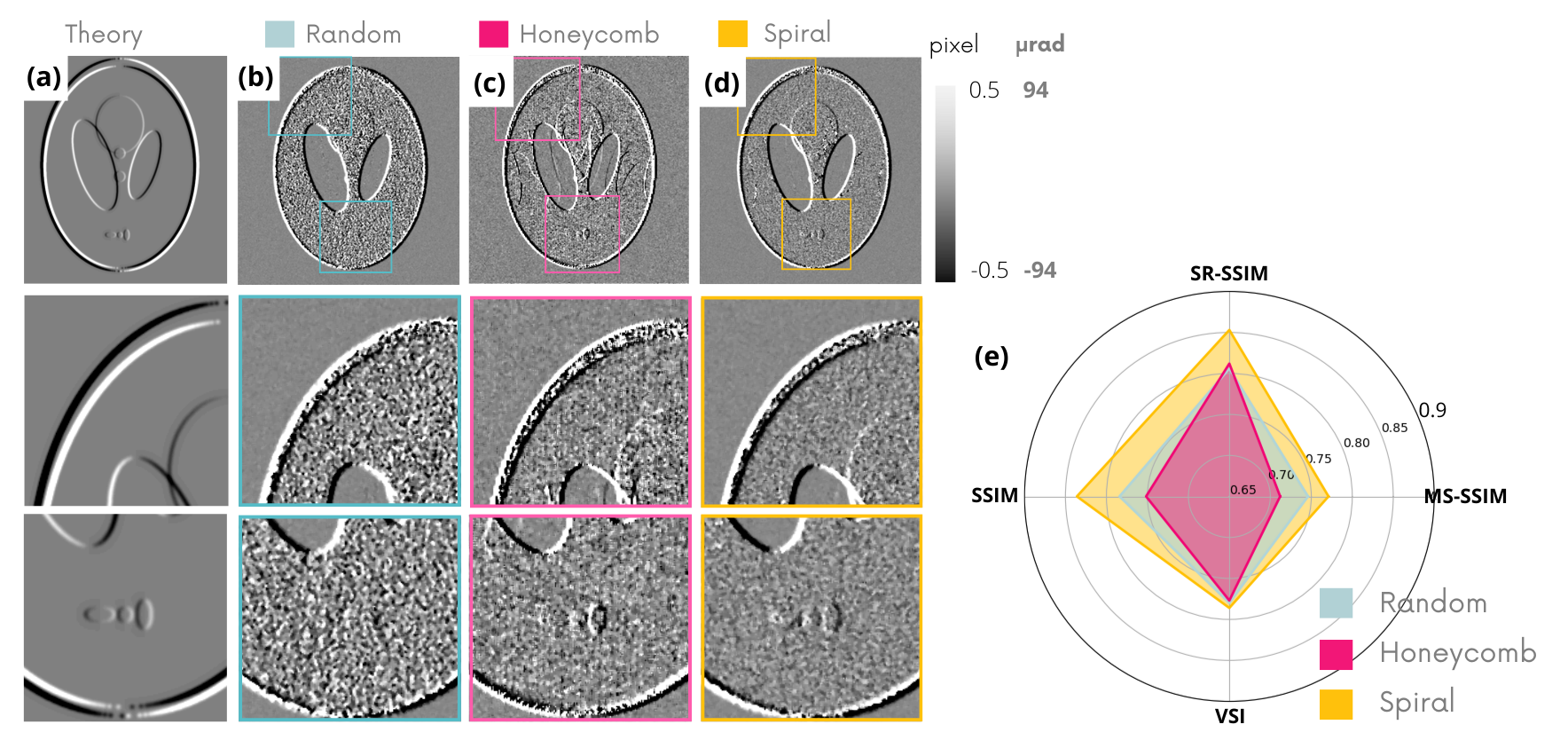}
\caption{$D_y$ images comparison on SL sample obtained in theory (a) and with the different membrane geometries (b,c,d). Associated image quality indexes (e)}
\label{SL}
\end{figure}

\begin{figure}[h!]
\centering
\includegraphics[scale=0.45]{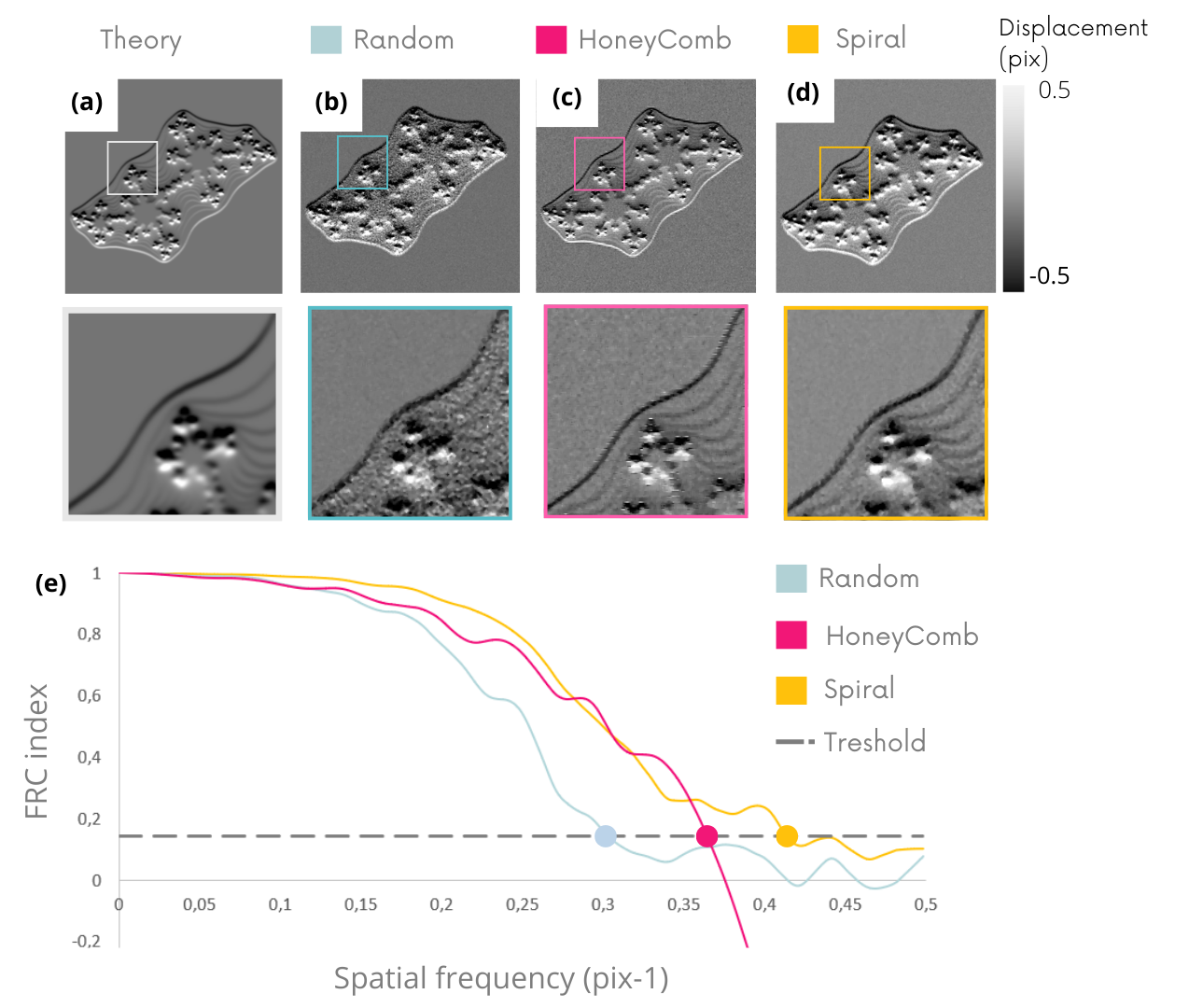}
\caption{Vertical refraction ($D_y$) images comparison on fractal sample obtained in theory (a) and with the different membrane geometries (b,c,d). FRC index computed on the 3 dy images (e) with a threshold at 1/7}
\label{FRC}
\end{figure}

NRMSE is known not to reflect the human eye vision for image quality \cite{wang2009mean}. Further image quality indices were considered in sub-figures \ref{indexes} (a,b,c,d) to validate the superiority of the spiral membrane, particularly in the case of unidirectional membrane displacement. The regular membrane was eliminated from the analysis due to low-quality results. The set of criteria compares the images obtained with theoretical reference images and the higher the metrics is the better the image quality. Figure \ref{indexes} (a,b,c,d) confirms the trends observed earlier on the NRMSE for 4 simplest object geometries. First of all, the random pattern seems to be the worst topology of the three tested even though it is the most used in literature. On the other hand, the spiral pattern is the best topology for the different geometries except the tore in which hexagonal and spiral are tied. The more complex Shepp Logan and fractal samples were analyzed in greater detail and are presented in Figure \ref{SL} and \ref{FRC}. 

Figure \ref{SL} (a,b,c,d) shows the refraction maps in the vertical direction ($D_y$)  obtained with the different membrane topologies for the Shepp-Logan phantom together with the theoretical image to retrieve. This phantom is made up of ellipsoids of different densities/thicknesses aiming to model a brain sample. This phantom, commonly used in biomedical imaging, emphasizes the strengths and weaknesses of different membrane geometries. Two different insets are isolated to highlight the background and signal-to-noise ratio. The subfigure \ref{SL} (e) shows in radar plot 4 different image quality metrics.
 
In this figure, the image obtained with the random pattern is the noisiest in the uniform thickness region and shows moderate noise in the background. The image retrieved with the hexagonal pattern contains long-line artifacts and a high level of noise in the background. Finally, the spiral pattern generates a low-noise refraction map with a higher level of detail. These qualitative observations are supported by Figure \ref{SL} (e), which shows that the spiral pattern outperforms the other geometries for all criteria. 

Table \ref{angular_s} lists the calculated angular sensitivity values of these images, as in \cite{Endrizzi}, carried in 8 different 5x40 pixel windows in a sample-free zone by calculating the mean and standard error of the standard deviation of the refraction angles. Once again, the spiral membrane achieves the best resolution with the lowest angular sensitivity down to 98 nano radians.

\begin{table}[h!]
\centering
\caption{Angular sensitivity in $D_y$ SL images}
\begin{tabular}{|c|c|c|c|}
    \hline
        & Random & Honeycomb & Spiral  \tabularnewline
    
    \hline
     Angular sensitivity (nrad)  & 141 & 810 & 98 \tabularnewline
    
    \hline
\end{tabular}
\label{angular_s}
\end{table}

Figure \ref{FRC} (a,b,c,d) shows the vertical refraction image ($D_y$) obtained on a fractal sample and Figure \ref{FRC} (e) the calculation of the Fourier ring correlation \cite{koho2019fourier} (FRC) associated with each of these images. Fourier Ring Correlation, widely used in microscopy, is another highly informative index to look at, relevant for analyzing a fractal object, since it has been established for calculating spatial resolutions.  
The FRC curves highlight the superiority of the spiral membrane, which achieves a resolution of 2.3 pixels compared with 2.7 and 3.2 pixels for the hexagonal and random membranes respectively. These spatial frequency estimates are consistent with visual impressions and the distribution of noise inside and outside of the sample in the $D_y$ images Figure \ref{FRC} (b,c,d). The results on the other component $D_x$ is equivalent.

\subsection{Experimental validation}
Previous numerical simulations showed the superiority of the spiral geometry. Experimental measurements were carried out to confirm its performance. To this end, 3 membranes were selected, manufactured and compared under experimental conditions as close as possible to the simulation parameters:
\begin{itemize}
    \item A piece of p180 sandpaper generates a random reference pattern.
    \item A honeycomb pattern of holes of 60 µm diameter in a 30-µm-thick Ni foil.
    \item A spiral pattern reproducing the Vogel pattern of holes of 60 µm diameter in a 30-µm-thick Ni foil.
\end{itemize}

A fractal object similar to the one used in simulation was printed in 3D and imaged with these different membranes. Ten pairs of reference and sample images were acquired, each with an exposure time of 30 seconds to limit the effects of photon noise. The membrane was moved  with a random displacement of between 50 and 150 pixels between each pairs in only one direction to get closer to numerical simulation conditions.

\begin{figure}[h!]
\centering
\includegraphics[scale=0.45]{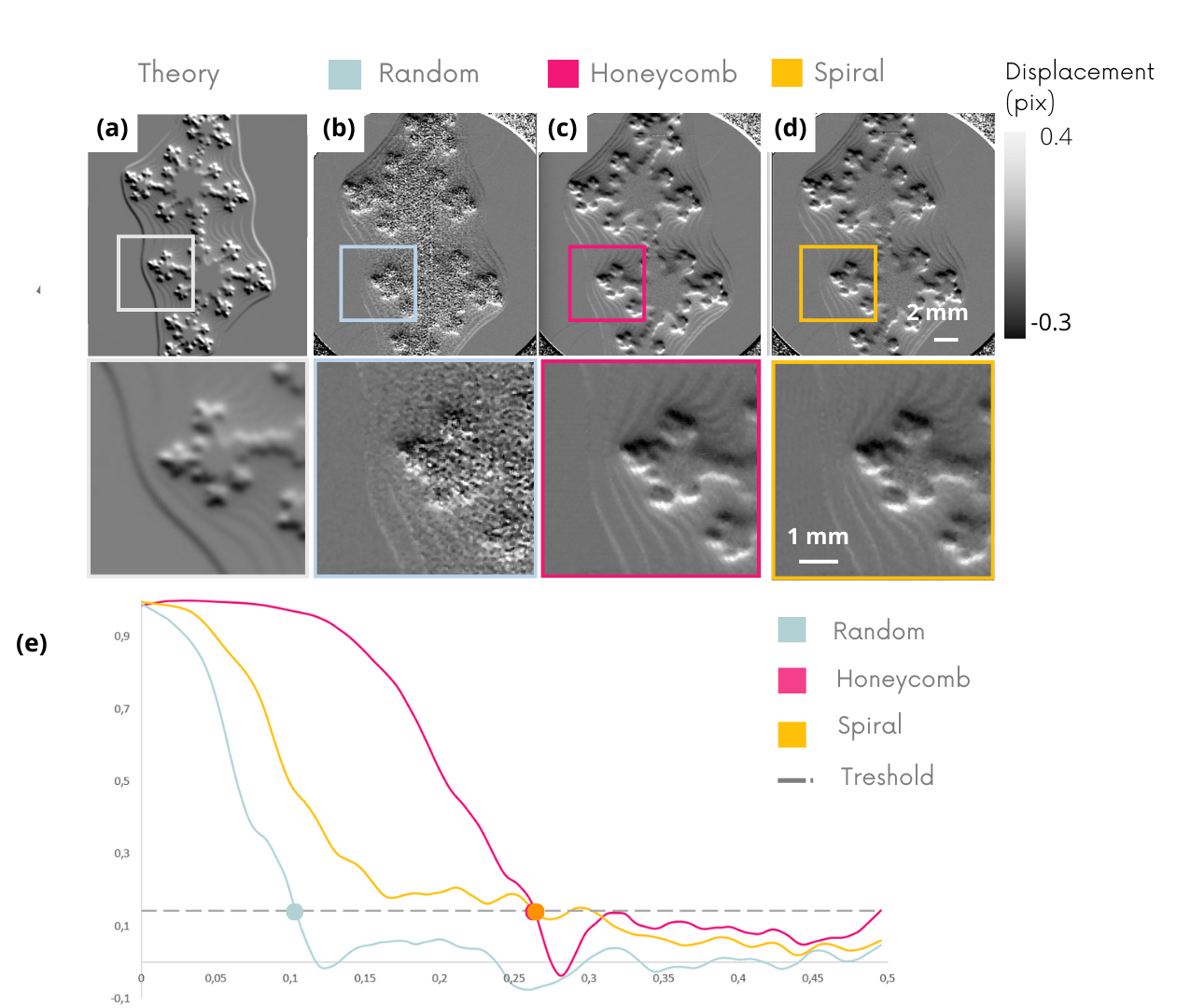}
\caption{Experimental validation. $D_y$ images comparison on fractal sample obtain in theory (a) and with the different membrane geometries (b,c,d). FRC index computed on the 3 $D_y$ images (e) with a threshold at 1/7.}
\label{experiments}
\end{figure}

Figure \ref{experiments} shows the vertical refraction ($D_y$) images obtained experimentally (b,c,d) with the different masks. The FRC curve, sub-Figure \ref{experiments} (e), shows the resolutions achieved.

This experimental analysis highlights the clear inferiority of random pattern compared with other geometries, confirming the previous numerical results. The image qualities achieved with the hexagonal and spiral patterns are close both visually and by analysis of the FRC curves.

\begin{figure}[h!]
\centering
\includegraphics[scale=0.35]{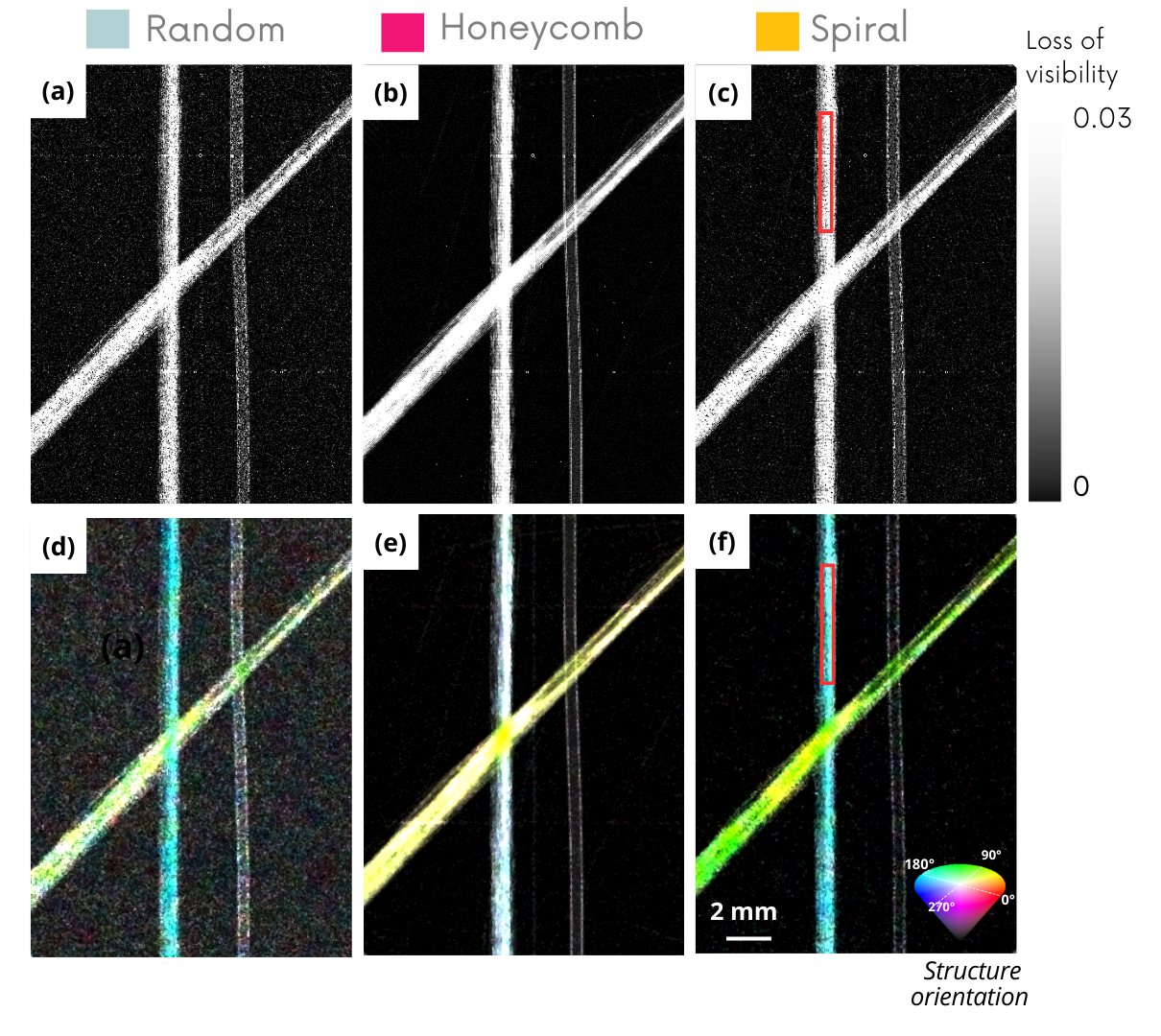}
\caption{Dark-field (top) and directional dark-field (bottom) images on a nylon wire and bundles of carbon fibers sample using 3 different membranes : random (a,d), hexagonal (b,e) and spiral (c,f) patterns}
\label{DF}
\end{figure}

To go further, and not restrict the study to the case of phase contrast, the experimental investigation was extended to the quality of dark-field and directional dark-field images.  Figure \ref{DF} shows the results of this comparison. 
Directional dark-field is color coded using the HSV  (Hue, Saturation, Value) space.  In details, hue is the angle of the semi major axis (modulao $\pi$), saturation is the length of this  principal vector of the dark field tensor, and finally the value encodes the norm of this tensor (i.e the square root of the squared axis lengths). As there is currently no reliable simulation model capable of faithfully reproducing the effects measured in dark-field imaging, this evaluation was carried out exclusively on an experimental approach.

The sample used for this purpose is a home-made phantom consisting in a nylon wire and two bundles of carbon fibers with two different orientations. Fiber bundles are known to scatter anisotropically, so they have a strong dark-field signal with a preferred orientation. In contrast, the nylon wire generates no scattering signal.

In Figure \ref{DF}, dark-field and directional dark-field images show once again the inferiority of sandpaper, which induces noisier dark-field images inside and outside of the sample. 
\begin{table}[h!]
\centering
\caption{Signal to noise ratio in dark-field images and saturation to noise ration in directionnal dark-field images in the ROI represented by the red boxes in figure \ref{DF}.}
\begin{tabular}{|c|c|c|c|}
    \hline
        & Random & Honeycomb & Spiral  \tabularnewline
    
    \hline
     SNR DF & 9,8 & 34 & 13,7 \tabularnewline
     \hline
     SNR DDF saturation  & 5 & 3 & 8 \tabularnewline
    
    \hline
\end{tabular}
\label{SNR}
\end{table}

Dark-field images show that the hexagonal pattern mask is the best candidate for reducing image noise. The spiral membrane gives inferior results to the hexagonal geometry, but significantly better than those of the random membrane, as reflected in the signal-to-noise ration (SNR) values in Table \ref{SNR}.
However, the directional dark-field images reveal the limitations of the honeycomb geometry : although it reduces image noise, the orientations of the structures are poorly evaluated. This visual observation translates into a lower saturation value, as shown in the last line of the Table \ref{SNR}. The spiral membrane therefore emerged as the best compromise, enabling us to improve image quality while producing directional dark-field images with sharper orientation values, with stronger coloration and better contrast.

\newpage
\section{Discussion}

In this article we estimated the image quality in X-ray phase contrast and dark-field imaging using different illumination patterns. This work was carried out by simulation and validated experimentally.

The overall numerical and experimental study demonstrates the value of membranes with quasi-periodic geometries, such as spiral patterns, for improving phase and dark-field image quality. This analysis highlights the superiority of spiral patterns over the other geometries considered conventionally: random, regular and honeycomb. The use of regular patterns such as grids creates significant artifacts due to the preferred directions contained within the pattern. Finally, the study shows that random patterns such as those obtained by the illumination of sandpapers used up to now are largely disadvantageous compared to other geometries.

The efficiency of the different patterns has been studied here empirically. A purely theoretical solution in terms of image reconstruction equations seems beyond reach. We evaluated in this study different criteria such as local extrema distribution, intensity gradient values, contrast, modulation distributions and packing. As the mask optimization problem is relatively complex, no single criterion can explain the above empirical observations. 

We propose to explain these results in terms of two key parameters: The intensity gradient distribution and  the modulation packing density obtained on reference images using the different masks.

\subsection{Gradient distribution}
Starting from equation (\ref{eq:TIE_LCS}), we can see that the resolution of phase images (i.e. of displacement $D_x$, $D_y$) are linked to the term $\nabla Ir$, the gradient of the reference image. 
One way of understanding the superiority of some geometries would be to think of these patterns as optimizing the distribution and gradient values of reference images, and thus improving the resolution of displacement terms $D_x$ and $D_y$.

\begin{figure}[h!]
\centering
\includegraphics[scale=0.23]{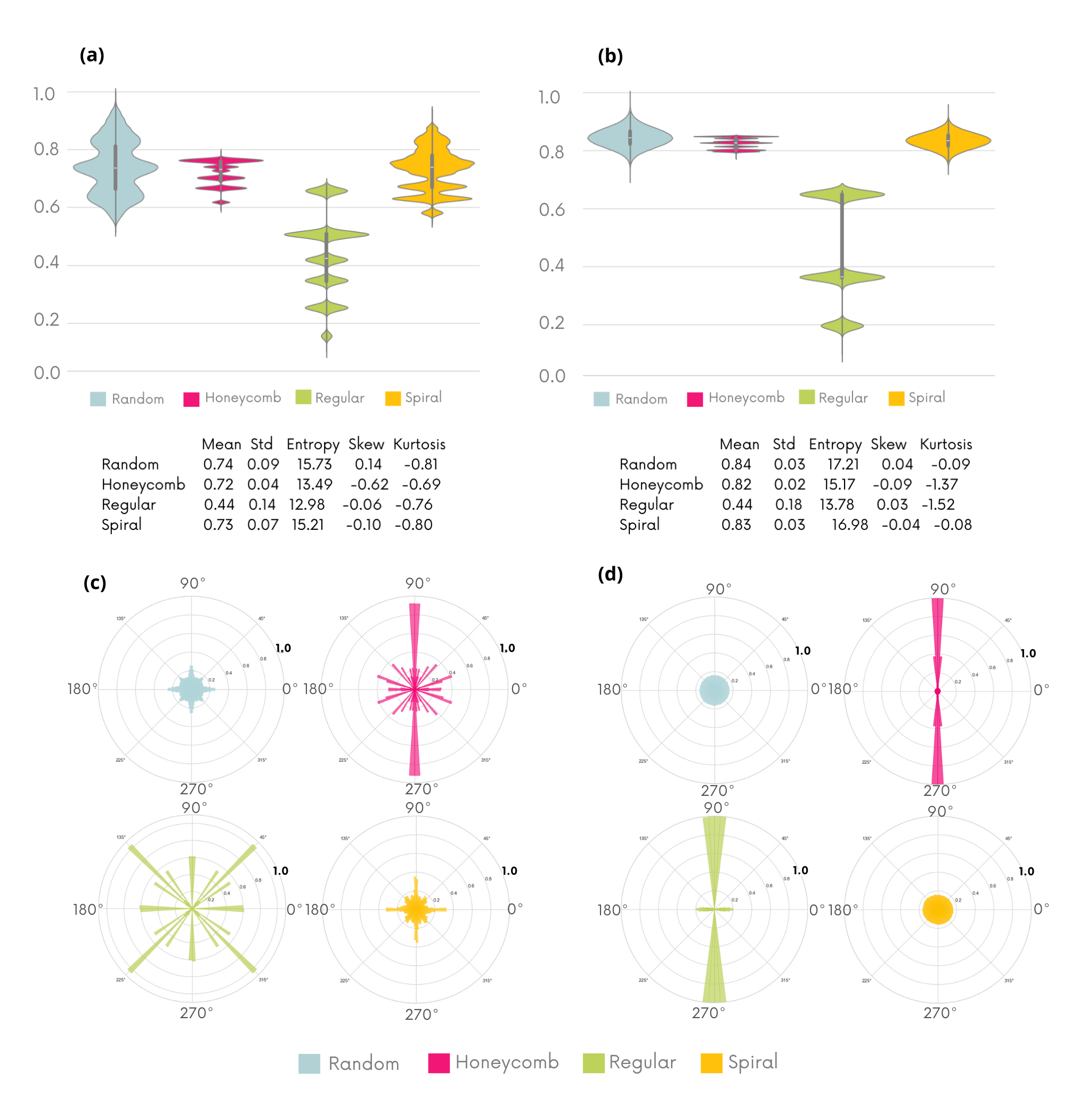}
\caption{Analysis of gradient distribution $\lVert \nabla(I_r) \rVert$ on a single reference image (a) and on the sum of 10 reference images (b). Gradient orientation distribution on a single reference image (c) and the sum of ten reference images (d).}
\label{grad}
\end{figure}

Figure \ref{grad} shows the intensity gradient distributions on the reference images alone (a) and on the sum of the 10 reference images used (b). Sub-Figures (c) and (d) show the distribution of gradient orientations on reference images alone, then on the sum of the 10 reference images.

Graphs (a,b) show that random and spiral geometries address a wide range of gradient values in a continuous way. In contrast, the distributions associated with hexagonal and regular patterns are discretized and more concentrated at the level of mean gradient values.
Sub-figures c and d show that for hexagonal and regular geometries, the gradients calculated on the reference images have clear preferred directions. This effect is reinforced on the sum of the 10 reference images, while for the random and spiral patterns the orientation distribution is homogeneous. This directionality in gradient distribution can be the source of geometric artifacts in reconstructed images.
This observations helps to explain the superiority of the spiral pattern over the hexagonal and regular geometries, but not the inferiority of the random distribution. An analysis of the modulation density gives some additional insight on the differences in image quality..

\subsection{Modulation packing density}

\begin{figure}[h!]
\centering
\includegraphics[scale=0.3]{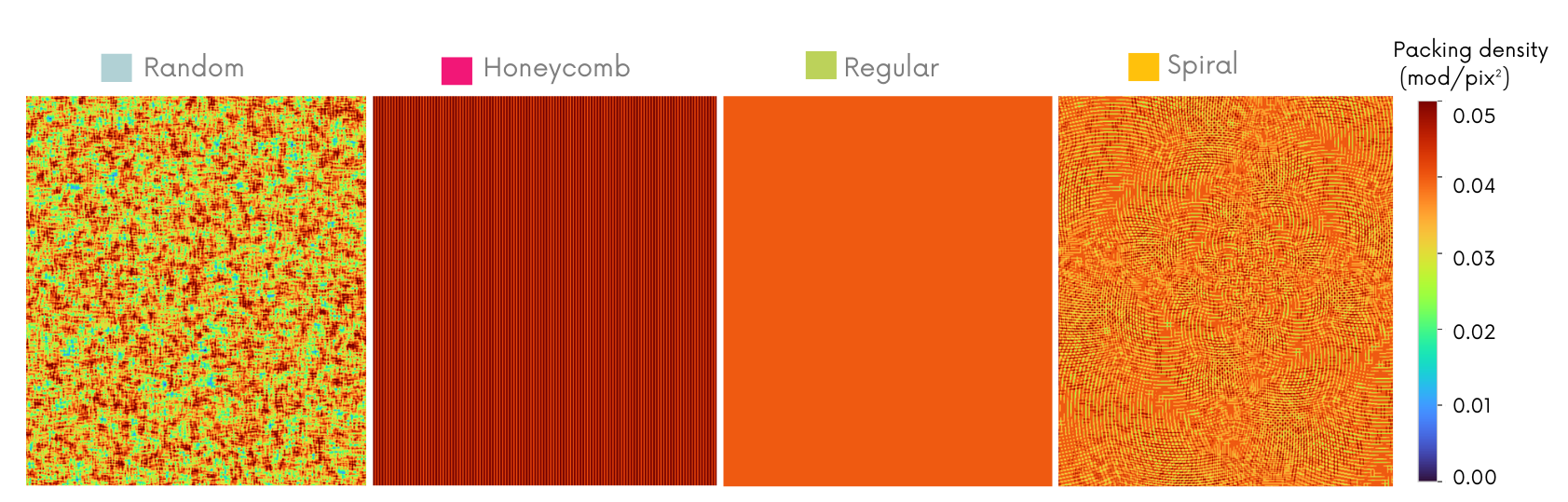}
\caption{Analysis of packing density distribution ($modulations/pixel^2$) on a single reference image.}
\label{packing}
\end{figure}

In order to explain the variations in image quality obtained with the different masks, the analysis of the distribution of intensity gradient values must be compared with the spatial density of modulation distribution. Indeed, the gradient values obtained on reference images may be ideal, but if the number of modulations contained in the latter is insufficient, the retrieved phase image will be of poor quality, due to the lack of information caused by the local absence of modulation. 
Figure \ref{packing} shows the modulation density distribution map (in $modulation/pixel^2$) on reference images obtained by simulation with the 4 mask geometries.

This figure shows the weakness of the random pattern with an average modulation density well below the others. Visually, the Figure \ref{packing} describes an inhomogeneous distribution of modulations on the reference image obtained with this pattern. Despite some areas of high density (red), the image surface contains many regions of insufficient modulation density (blue and green), which can lead to information loss and artifacts when reconstructing phase and dark-field data.
In contrast, regular and hexagonal patterns maximize and homogenize modulation distribution. The spiral pattern achieves an average modulation density comparable to that of the regular pattern, while evenly distributing the modulations over the image surface, making it clearly superior to the random pattern in this respect.

Finally, we can explain the advantages of the spiral pattern in phase contrast image enhancement, as it allows us to optimally pave the space (without having a preferred direction, as can be the case with regular structures which cause artifacts) while maximizing the gradient values addressed by the modulation.
This analysis also helps us to understand the weaknesses of the random pattern, which can achieve interesting gradient values, but whose paving is often locally insufficient, resulting in a loss of information during phase retrieval.


The results highlight that the effectiveness of modulation geometries depends not only on their ability to create appropriate intensity gradients, but also on their spatial distribution. While random patterns offer uniform intensity gradient distributions, their inhomogeneous spatial distribution severely compromises the quality of reconstructed images. Conversely, spiral patterns, with their ability to evenly distribute modulations over the imaging field, offer an optimal balance between gradient and density, resulting in images of higher for a given number of image pairs.
In addition, the use of quasi-periodic geometric patterns minimizes the directional artifacts that can be observed when using regular or hexagonal masks. 

This qualitative study provides an initial overview of the impact of the performance of different geometries on the quality of X-ray images reconstructed using the single mask method. This work can be taken further in various ways, by exploring other geometric configurations or developing theoretical models to improve our understanding of the impact of membrane structure on image quality. Such work could lead to further optimization of the masks used and the quality of the reconstructed images. Moreover this optimization was performed on X-ray single mask imaging but other techniques using similar structured illumination scheme, such as visible light microscopy \cite{gigan2022imaging, boominathan2020phlatcam} or wavefront sensing \cite{berto2017wavefront} might benefit of this study. In this study, we focused on the topology of the membrane using several pairs of randomly scanned images. In the near future, we will work on determining whether this optimization is also valid in single-shot approaches \cite{qiao2021single,doherty2023single} or if we optimize the acquisition scheme for a given membrane.

\subsection{Data Availability}
All data were made publicly available after acceptance of the article. The synthetic data were generated using Paresis \cite{Paresis} and will be made accessible on open open-access repository including the processed image. The same repository will include the experimental data and the phase retrieved image.

\section{Acknowledgments}
Part of this work was supported by the LABEX PRIMES (ANR-11-LABX-0063) of Université de Lyon, within the program  "Investissements d’Avenir"  (ANR-11-IDEX-0007) operated by the French National Research Agency (ANR). We acknowledge the support of ANRT cifre n°2022/0476 program that partly funds C.M.. E.B acknowledges the support of Inserm IRP linx project.

\section*{Author contributions}
The authors contributed equally to all aspects of the article. 

\section*{Competing interests}
The authors declare no competing interests except that C.M, D.M.C, B.L, B.F are or were employees of Xenocs.


\bibliography{sample}

\end{document}